\documentclass[prd,twocolumn,floatfix,preprintnumbers,letterpaper]{revtex4}

\usepackage{graphicx}
\input{epsf}
\usepackage{epsf}
\usepackage{graphicx,epsfig}
\usepackage{bm}
\usepackage{latexsym,amssymb,amsmath,float}

\newcommand {\ga} {\ {\raise-.5ex\hbox{$\buildrel>\over\sim$}}\ }
\newcommand {\la} {\ {\raise-.5ex\hbox{$\buildrel<\over\sim$}}\ }
\newcommand {\La}{\Lambda}
\newcommand{\be}{\begin{equation}}
\newcommand{\ee}{\end{equation}}
\newcommand{\bea}{\begin{eqnarray}}
\newcommand{\eea}{\end{eqnarray}}

\begin{document}

\title{Anapole Dark Matter}
\author{Chiu Man Ho and Robert J. Scherrer}
\affiliation{Department of Physics and Astronomy, Vanderbilt University,
Nashville, TN  37235}

\begin{abstract}
We consider dark matter (DM) that interacts with ordinary matter exclusively
through an electromagnetic anapole, which is the only allowed
electromagnetic form factor for Majorana fermions.  We show that unlike
DM particles with an electric or magnetic dipole moment, anapole
dark matter particles annihilate exclusively into fermions via purely $p$-wave
interactions, while tree-level annihilations into photons are forbidden. We calculate the
anapole moment needed to produce a thermal relic abundance in agreement with cosmological
observations, and show that it is consistent with current XENON100 detection limits
on the DM-nucleus cross-section for all masses, while lying just below the detection
threshold for a mass $\sim 30-40$ GeV.
\end{abstract}

\maketitle

\section{Introduction}
\label{sec:introduction}

Roughly $20-25$\% of the total energy content of the universe is in the form of
non-baryonic dark matter (DM).  The exact nature of the dark matter remains a
mystery.  Recently, a variety of authors have explored the possibility that
the dark matter might interact electromagnetically with ordinary matter, via
an electric or magnetic dipole moment \cite{Pospelov,Sigurdson,Gardner,Masso,Fitzpatrick,Cho,Heo1,Heo2,Banks,Barger1,Fortin,
Nobile,Barger2,Heo3}.  Direct detection experiments strongly constrain such dipole
moments for particle masses $\ga 10$ GeV. In particular, electric and magnetic dipole moments sufficiently
small to evade direct detection limits cannot provide the correct thermal relic abundance for the dark
matter unless $m \la 10$ GeV \cite{Sigurdson,Masso,Fortin}.

Although most of the interest in electromagnetic form factors has been concentrated on the electric and
magnetic dipole moments, Pospelov and ter Veldhuis \cite{Pospelov} considered several other possible
forms of electromagnetic coupling to the dark matter, including the electric quadrupole moment and the anapole moment.
It is the latter which we consider in more detail here. The anapole moment was first proposed by
Zel'dovich \cite{Zeldovich}.  Unlike the electric and magnetic dipole moments, the anapole moment has no
classical analog, as it does not correspond to a multipolar distribution. It is related to the toroidal dipole moment,
which corresponds to a solenoid with the ends joined into a torus, producing an azimuthal magnetic field (for an explanation
of the difference between the anapole moment and the toroidal dipole moment, see, e.g., Ref. \cite{Dubovik}).
The first experimental measurement of an anapole moment in atomic nuclei
was noted by Woods et al. \cite{Wood}.

Here, we consider a DM particle that interacts with ordinary matter entirely through an
electromagnetic anapole moment.  For Majorana fermions, the anapole is the only allowed
electromagnetic form factor, making it particularly interesting.
Anapole dark matter coupling to a dark photon was previously considered by Fitzpatrick and Zurek \cite{Zurek}, but our model
differs from theirs in that we have direct coupling to the Standard Model (SM) photons, with a
correspondingly different Lagrangian.
In the next section, we begin with the interaction Lagrangian for anapole dark matter and explain the
general properties of the anapole moment. In Sec. \ref{sec:relic}, we use the anapole moment operator to
derive the total annihilation cross-section.  We show that it leads, at tree-level, to
pure $p$-wave annihilation into fermions.
%In Sec., we
%derive a new lower bound on the mass of a thermal dark matter particle from CMB limits on the number
%of effective neutrino degrees of freedom.
We then use this total
annihilation cross-section to calculate the relic density of the anapole dark matter. In Sec. \ref{sec:direct},
we calculate the cross-section for scattering of the anapole dark matter particle
off of nuclei, and compare this result to the recent XENON100 limits.
We find that, unlike the case of
the electric or magnetic dipole, anapole dark matter with an arbitrary mass
can provide the correct thermal relic abundance for
dark matter and evade current direct detection limits.
Our results are discussed in Sec. \ref{sec:discuss}.

\section{Dark Matter Anapole Moment}
\label{anapole}

A Majorana fermion, by definition, is a CPT self-conjugate particle.
Since the interaction energies for both electric and magnetic dipole moments are CPT-odd, a Majorana fermion cannot
acquire any of these dipole moments (although transition electric and magnetic dipole moments are still possible \cite{Masso}). In fact, it has been proven that for a massive spin-$S$ Majorana fermion, the only electromagnetic form factors allowed are the anapole moment and higher multipoles of it \cite{MajoranaFermion}. The maximum number of these multipoles is $2\,S$. This theorem has been generalized and shown to be valid for any
CPT self-conjugate massive particle with spin-$S$, be it fermionic or bosonic \cite{MajoranaParticles}. For CPT self-conjugate massless particles, except the spin-half fermions, no electromagnetic couplings are allowed.

In this article, we consider the dark matter to be a Majorana fermion with spin $1/2$. According to the theorem mentioned above, the only electromagnetic form factor allowed is the anapole moment. The interaction operator for this anapole moment is of the form $\bar{\chi}\,\gamma^\mu\, \gamma^5\, \chi\, \partial^\nu  F_{\mu\nu}$ where $\chi$ is the dark matter and $F_{\mu\nu}$ is the electromagnetic field strength tensor.
(We remark that the corresponding anapole interaction operator for a Dirac fermion takes exactly the same form.)
Since this is a dimension-6 operator, the interaction Lagrangian density is given by
\bea
\label{lagrangian}
\mathcal{L}_{I} = \frac{g}{\Lambda^2}\,\bar{\chi}\,\gamma^\mu\, \gamma^5\, \chi\, \partial^\nu F_{\mu\nu}\,,
\eea
where $g$ is the coupling constant and $\La$ is the cut-off scale. This
interaction operator breaks charge conjugation symmetry C and parity symmetry P,
but is invariant under time-reversal symmetry T. In contrast to the electric and
magnetic dipole moments which interact with external electromagnetic fields, the
anapole moment has the unique feature that it interacts only with external electromagnetic currents $J_\mu = \partial^\nu F_{\mu\nu}$.

The interaction operator for the anapole moment is related to but different from that for the toroidal
dipole moment, which has the form $\bar{\chi}_1\,\gamma^\mu\, \gamma^5\, \chi_2\, \partial^\nu
F_{\mu\nu}$. One can visualize the idea of the toroidal dipole moment as
follows: a solenoid is
folded to join the ends into a torus and so the current configuration is such that an azimuthal magnetic
field is generated. The toroidal dipole vector then points in the direction dictated by the curl of the
magnetic field. Obviously, the toroidal dipole moment reduces to the anapole moment in the limit
$\chi_1=\chi_2$ \cite{Dubovik}. In other words, when the incoming and outgoing particles are the same,
the toroidal dipole moment coincides with the anapole moment. This means that we can visualize the idea
of the anapole moment in a similar way as that of the toroidal dipole moment.
For instance, if neutrinos are Majorana in nature, they can acquire both a toroidal dipole moment and an
anapole moment formed from the various neutrino fields in the mass basis. It has been shown that if
neutrinos (Dirac or Majorana) indeed have a toroidal dipole moment, it will lead to transition radiation
when the neutrino crosses the interface between two media \cite{transition_rad}.

%%%%%%%%%%%%%%%%%%%%%%%%%%%%%%%%%%%%%%%%%%%%%%%%%%%%%%%%%%%%%%%%%%%%%%%%%%%%%%%%%%%%%%%%%%%%%%%%%%%%%%%%%%%%%%%%%%%%%%%%%%%%%%%%%%%%%%%%
\begin{table}[t!]
  \centering
  \begin{tabular}{l*{4}{c}r}
             ~~~~~  & $\vec{\sigma}$  & ~~~~ $\vec{E}$  & ~~~~ $\vec{B}$  & ~~~~ $\vec{J}$ \\ \hline \hline
C            ~~~~~  & +               & ~~~~ --         & ~~~~ --         & ~~~~ --   \\ \hline
P            ~~~~~  & +               & ~~~~ --         & ~~~~ +          & ~~~~ --   \\ \hline
T            ~~~~~  & --              & ~~~~ +          & ~~~~ --         & ~~~~ --   \\ \hline
CPT          ~~~~~  & --              & ~~~~ +          & ~~~~ +          & ~~~~ --   \\ \hline
\end{tabular}
  \caption{Transformation properties of the interaction energies for electric dipole moment, magnetic dipole moment and anapole moment under C, P, T.}
\label{table}
\end{table}
%%%%%%%%%%%%%%%%%%%%%%%%%%%%%%%%%%%%%%%%%%%%%%%%%%%%%%%%%%%%%%%%%%%%%%%%%%%%%%%%%%%%%%%%%%%%%%%%%%%%%%%%%%%%%%%%%%%%%%%%%%%%%%%%%%%%%%%%%

The transformation property of the anapole moment operator under C, P, T\, is
perhaps more transparent in the non-relativistic limit, at which the interaction energy takes the form
%$\frac{g}{\Lambda^2}\,(\vec{\sigma}\,\cdot\,\vec{J})$,
\bea
\mathcal{H}_{I} = -\frac{g}{\Lambda^2}\,\vec{\sigma}\,\cdot\,\vec{J}\,,
\eea
where $\vec{\sigma}$ are the Pauli spin matrices and $\vec{J}=\vec{\nabla}\times
\vec{B}$ is the electromagnetic current. This is consistent with the intuitive
picture we described above by visualizing the folded solenoid, namely we need a
non-zero $\vec{\nabla}\times \vec{B}$ to generate an anapole moment. In Table
I, we compare the transformation properties of the interaction energies for the electric dipole moment ($\vec{\sigma}\,\cdot\, \vec{E}$), magnetic dipole moment ($\vec{\sigma}\,\cdot\, \vec{B}$) and anapole moment ($\vec{\sigma}\,\cdot\,\vec{J}$) under C, P, T \,in the non-relativistic limit.
As one can easily read off from Table I, the interaction energy for the anapole moment violates C and P individually but preserves T. This is consistent with the transformation properties of the anapole interaction operator itself in \eqref{lagrangian} under C, P, T.

A similar but different anapole interaction operator has been considered by \cite{Zurek} and takes the form $\bar{\chi}\,\gamma^\mu\, \gamma^5\, \chi\, A'_\mu$. If we assume that $A'_\mu$ is the SM photon, then in order to maintain gauge invariance, the field $\chi$ is required to transform as
$\chi \rightarrow e^{-i\,\gamma^5\,\xi}\,\chi$ simultaneously when the photon transforms as $A'_\mu \rightarrow A'_\mu -\partial_\mu \xi$.
Indeed, the kinetic operator plus the interaction operator, $\bar{\chi}\,i\,\gamma^\mu\,\partial_\mu\,\chi + \bar{\chi}\,\gamma^\mu\, \gamma^5\, \chi\, A'_\mu $, is invariant under the gauge transformations: $A'_\mu \rightarrow A'_\mu -\partial_\mu \xi$
and $\chi \rightarrow e^{-i\,\gamma^5\,\xi}\,\chi$.
So the field $\chi$ acquires a chiral symmetry.
However, this is impossible unless $\chi$ is massless
because the mass term $m_\chi\,\bar{\chi}\,\chi$ breaks the chiral symmetry.
Since a dark matter particle must be massive, $A'_\mu$ cannot be the SM photon. Nevertheless, it could still be possible that $A'_\mu$ represents a dark photon which kinetically mixes with the SM photon through the operator $\epsilon \,F^{\mu\nu}\,F'_{\mu\nu}$ \cite{Holdom,Nima,Essig,Pospelov2}.

\begin{figure}[t!]
%\centerline{\epsfxsize=2.5truein\epsffile{annihilationFF.eps}}
\includegraphics[height=3cm, width=6cm]{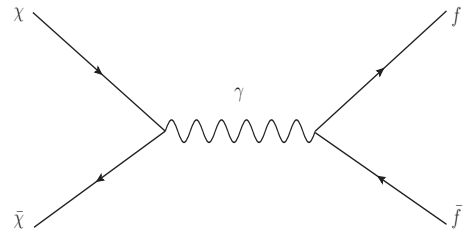}
\caption{$\chi\,\bar{\chi} \rightarrow f\,\bar{f}$.}
\label{annihilationFF}
\end{figure}

\begin{figure}[t!]
%\centerline{\epsfxsize=2.5truein\epsffile{annihilationPhotons1.eps}}
\includegraphics[height=3cm, width=5cm]{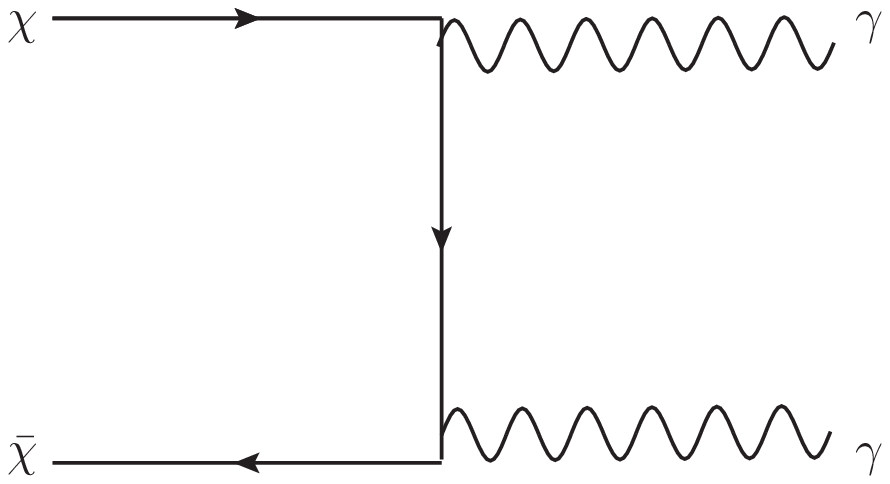}
\caption{t-channel $\chi\,\bar{\chi} \rightarrow \gamma \, \gamma$.}
\label{annihilationPhotons1}
\end{figure}

\begin{figure}[t!]
%\centerline{\epsfxsize=2.5truein\epsffile{annihilationPhotons2.eps}}
\includegraphics[height=3cm, width=5cm]{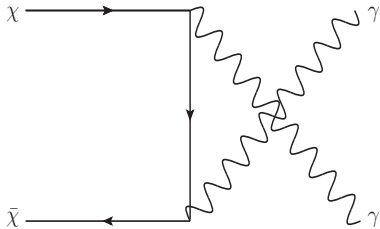}
\caption{u-channel $\chi\,\bar{\chi} \rightarrow \gamma \, \gamma$.}
\label{annihilationPhotons2}
\end{figure}

Therefore, for a fermionic dark matter to couple directly to SM photons, the interaction Lagrangian density in \eqref{lagrangian} gives the unique interaction operator for the anapole moment. At tree-level, this
operator allows for the annihilation process $\chi\,\bar{\chi} \rightarrow
f\,\bar{f}$ where $f$ is a kinematically allowed SM fermion. (See the Feynman
diagram in Fig. \ref{annihilationFF}.)
If we assume that the mass of the dark matter $m_\chi$ is smaller than
$M_W$, then the process $\chi\,\bar{\chi} \rightarrow W^{+}\,W^{-}$ is not
kinematically allowed. The process $\chi\,\bar{\chi} \rightarrow \gamma \,
\gamma$ is kinematically allowed but forbidden. (See, for instance, the Feynman diagrams
in Fig. \ref{annihilationPhotons1} and Fig. \ref{annihilationPhotons2}.
The two similar diagrams with crossed fermion lines for Majorana dark matter are not shown.)
The reason for $\chi\,\bar{\chi} \rightarrow \gamma \,
\gamma$ being forbidden at tree-level
is that $\partial^\nu F_{\mu\nu} = \partial_\mu\, (\partial^\nu A_{\nu}) -
\partial^2 A_{\mu}$, and for on-shell photons, both $\partial^\nu A_{\nu}$ and
$\partial^2 A_{\mu}$ are zero. A more intuitive way to understand this fact is
as follows. The anapole dark matter only couples to the external electromagnetic
current that generates the electromagnetic fields. But the on-shell external
photons do not constitute such an electromagnetic current.
This is in sharp contrast to dark matter with electric and magnetic dipole
moments, both of which allow for the process $\chi\,\bar{\chi} \rightarrow \gamma \, \gamma$ at tree-level.

%%%%%%%%%%%%%%%%%%%%%%%%%%%%%%%%%%%%%%%%%%%%%%%%%%%%%%%%%%%%%%%%%%%%%
%%%%%%%%%%%%%%%%%%%%%%%%%%%%%%%%%%%%%%%%%%%%%%%%%%%%%%%%%%%%%%%%%%%%%

%%%%%%%%%%%%%%%%%%%%%%%%%%%%%%%%%%%%%%%%%%%%%%%%%%%%%%%%%%%%%%%%%%%%%
%%%%%%%%%%%%%%%%%%%%%%%%%%%%%%%%%%%%%%%%%%%%%%%%%%%%%%%%%%%%%%%%%%%%%%%

%%%%%%%%%%%%%%%%%%%%%%%%%%%%%%%%%%%%%%%%%%%%%%%%%%%%%%%%%%%%%%%%%%%%%%
%%%%%%%%%%%%%%%%%%%%%%%%%%%%%%%%%%%%%%%%%%%%%%%%%%%%%%%%%%%%%%%%%%%%%%

\section{Cosmological relic abundance of anapole dark matter}
\label{sec:relic}

As mentioned in the previous section, for $m_\chi < M_W$, the only annihilation
channel for anapole dark matter is
$\chi\,\bar{\chi} \rightarrow f\,\bar{f}$ where $f$ is a kinematically allowed SM fermion. The thermally averaged annihilation
cross-section for this channel is found to be purely $p$-wave (see Appendix A
for the details of the calculation):
\bea
\langle\,\sigma_{\chi\,\bar{\chi} \rightarrow
f\,\bar{f}}\;v_{\textrm{rel}}\,\rangle=
\frac{4\,g^2\,\alpha\,m_\chi^2}{\La^4}\left(\frac{T}{m_\chi}\right)\,,
\eea
where $\alpha = e^2/4\pi \approx 1/137$ (note this convention differs
from that in Ref. \cite{Pospelov}) and $T$ is the temperature.

Using this annihilation cross-section and the methodology of Refs.
\cite{ST,KT}, we now derive the thermal relic abundance of $\chi$.  (See also
the more recent discussion in Refs. \cite{Steigman,Bender}). We will confine our attention to
masses in the range 10 ${\rm MeV} < m_{\chi} < 80 ~{\rm GeV}$.  The lower bound satisfies
the requirement that the annihilations of the dark matter particles as they become
nonrelativistic in the early universe not heat the
photons relative to the neutrinos and thereby violate CMB observations constraining the
number of relativistic degrees of freedom \cite{HoScherrer}.  The upper bound comes from the requirement that
$m_\chi < M_{W}$.  Although larger masses are not necessarily excluded, we choose to explore that possibility
elsewhere.

\begin{figure}[htb]
\centerline{\epsfxsize=3.7truein\epsffile{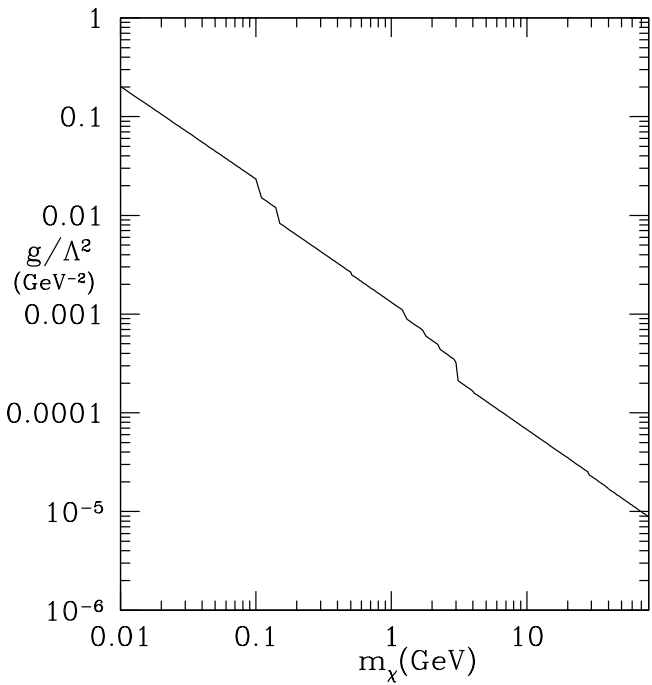}}
\caption{The anapole moment, $g/\Lambda^2$, as a function of the mass, $m_\chi$, needed
to account for the observed relic abundance of dark matter.}
\label{LambdaVSmass}
\end{figure}

We ignore the possibility of an asymmetry
between particles and antiparticles and also neglect the possibility of coannihilations \cite{Griest}.
Since we have a pure $p$-wave annihilation, we can write the total cross-section as
\begin{equation}
\sum_{m_f < m_\chi}\,\langle\,\sigma_{\chi\,\bar{\chi} \rightarrow f\,\bar{f}}\;v_{\textrm{rel}}\,\rangle = \sigma_0 \left(\frac{T}{m_\chi}\right),
\end{equation}
where $\sigma_0$ is given by
\begin{equation}
\label{sigma0}
\sigma_0 = \frac{4\,g^2\,\alpha\,m_\chi^2}{\La^4}\, N_f,
\end{equation}
and $N_f$ counts the effective number of annihilation channels with mass $m_f < m_\chi$. For each annihilation channel,
the contribution to $N_f$ is given by the square of the corresponding fermion charge ($Q^2$)
multiplied by the color factor whenever applicable.
For our mass range of interest, $N_f$ can
range from $N_f = 1$ (for $\chi \,\bar \chi \rightarrow e^+e^-$ only, \, if
$m_\chi \la 100$ MeV), up to $N_f = 20/3$ for $m_\chi > m_b$ (annihilation
into 3 charged leptons and 5 quark flavors, with each of the latter given by $3\, Q^2$ for the color factor and charge).
Following Refs. \cite{KT,ST} and assuming pure $p$-wave annihilation, we can
write the contribution of the anapole dark matter to the density in the form
\begin{equation}
\label{Omega}
\Omega_{\chi}\, h^2 = (\,2.14 \times 10^9\,)\, \frac{x_f^{2}\; ({\rm GeV})^{-1}}{g_*^{1/2}\, M_{Pl}\,
\sigma_0}.
\end{equation} \\
This equation is valid as long as $\chi$ drops out of thermal equilibrium before $e^+ e^-$ annihilation, which is clearly
the case for $m_\chi > 10$ MeV.
In Eq. (\ref{Omega}), $\Omega_{\chi}$ is the density of $\chi$ relative to the critical density,
$h$ is the Hubble parameter in units of 100 km sec$^{-1}$ Mpc$^{-1}$, $g_*$ is the number of
relativistic degrees of freedom in the universe when $\chi$ drops out of thermal equilibrium,
$M_{Pl}$ is the Planck mass, and $x_f$ is given by \cite{ST,KT}
%\begin{widetext}
\bea
\label{xf}
x_f &=& \ln \left[\,0.076\, \left(\,\frac{g_\chi}{g_*^{1/2}}\,\right)\, M_{Pl}\, m_\chi\, \sigma_0\,\right] \nonumber \\
&& ~ - \frac{3}{2}\, \ln\, \ln\, \left[\,0.076\, \left(\,\frac{g_\chi}{g_*^{1/2}}\,\right) \,M_{Pl}\, m_\chi\, \sigma_0\,\right]\,,
\eea
%\end{widetext}
with $g_\chi = 2$ being the internal degrees of freedom for the Majorana $\chi \,\bar{\chi}$ pair.
If we assume that $\chi\, \bar \chi$ accounts for all of the
dark matter, then we can substitute the observed value of $\Omega_{\textrm{DM}}\, h^2 = 0.11$ \cite
{WMAP7B} into Eq. (\ref{Omega}) and use Eqs. (\ref{Omega}) and (\ref{xf}) to solve for
$\sigma_0$ as a function of $m_\chi$.  Then Eq. (\ref{sigma0}) gives the anapole moment,\,
$g/\Lambda^2$, \,as a function of $m_\chi$.  This is plotted in Fig. \ref{LambdaVSmass}.  Note that the step
discontinuities are an artifact of the approximation used here (in which $g_*$ changes
sharply as a function of temperature and $N_f$ changes sharply as a function of mass).  Nonetheless, the results shown
in Fig. \ref{LambdaVSmass} would not change significantly with a more detailed calculation.

In any sensible model, we require $g \la 1$, and $m_\chi < \Lambda$.  This is the
case for the range of masses displayed in Fig. \ref{LambdaVSmass}.  If we set $g=1$ for naturalness, then
as $m_\chi$ is varied from 10 MeV to 80 GeV, $\Lambda$ varies from 2.2 GeV to 340 GeV. Here, $\Lambda$ characterizes the
scale above which a new physics, presumably a new particle, should exist.
This new particle, which acts as the mediator, must couple to the photon to generate the anapole operator.
We remark that for masses near the lower end of
our range of interest ($m\chi \sim 10$ MeV and $\Lambda$ on the order of a few GeV),
it is possible that existing experimental data already rule out this model, and we are currently
investigating this question.  However, near the upper end of the mass range considered here,
where $\Lambda$ is several hundred GeV, it is likely that any additional new physics would have thus
far escaped detection.

Finally, while it is not shown in Fig. \ref{LambdaVSmass}, we remark that anapole
dark matter with mass in the range  $m_\chi \lesssim$ 10 MeV could also generate
the correct relic abundance. As we have noted, $m_\chi \la 5-10$ MeV is excluded by CMB measurements
in the standard cosmological scenario \cite{HoScherrer}, but smaller masses
are possible if the cosmology is modified \cite{HSinprep}.

\section{Direct detection limits}
\label{sec:direct}

In this section, we consider the direct detection limits on the DM-nucleus scattering cross-section. The Feynman diagram
for this scattering process is shown in Fig. \ref{scatteringNucleus}. According to the derivation in Appendix B, this scattering cross-section
is found to be
\bea
\sigma =\frac{1}{2\,\pi}\left(\frac{g}{\La^2}\right)^2\,Z^2\,e^2\, M_{\chi N}^2\,v^2
\,\left(\,1+\frac{2\,M_{\chi N}^2}{m_N^2}\,\right)\,,
\eea
where $m_N$ and $Z$ are the nuclear mass and charge, $v$ is the velocity of $\chi$
in the lab frame (i.e., the nuclear rest frame) and
$M_{\chi N} = m_\chi\,m_N/(m_\chi+m_N)$ is the reduced mass of the $\chi$-nucleus system. A
similar expression for $\sigma$ has been previously
computed in \cite{Pospelov}, but
our result differs from that expression by a factor of
$1+ 2\,M_{\chi N}^2/ m_N^2$. However, if $m_\chi \ll
m_N$, then $M_{\chi N} \sim m_\chi$ and
$2\,M_{\chi N}^2/ m_N^2 \ll 1$ reducing our expression to the result
of Ref. \cite{Pospelov}. On the other hand, when $m_\chi \sim m_N$ or $m_\chi \gg m_N$, the factor
$2\,M_{\chi N}^2/ m_N^2$ could become significant.  In limit where $m_\chi \gg m_N$, we have
$1+ 2\,M_{\chi N}^2/ m_N^2 \rightarrow 3$.

%%%%%%%%%%%%%%%%%%%%%%%%%%%%%%%%%%%%%%%%%%%%%%%%%%%%%%%%%%%%%%%%%%
\begin{figure}[h!]
%\centerline{\epsfxsize=2.5truein\epsffile{scatteringNucleus.eps}}
\includegraphics[height=3cm, width=6cm]{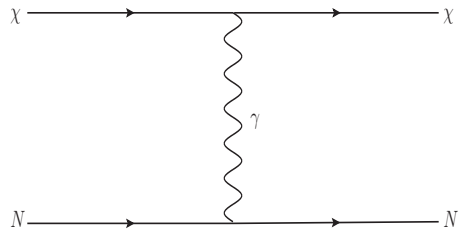}
\caption{$\chi\,N \rightarrow \chi\,N$.}
\label{scatteringNucleus}
\end{figure}
%%%%%%%%%%%%%%%%%%%%%%%%%%%%%%%%%%%%%%%%%%%%%%%%%%%%%%%%%%%%%%%%%%

A more useful quantity for the direct detection experiments is
\begin{widetext}
\be
\label{dsigma}
\frac{d\sigma}{dE_R} = \frac{1}{2\,\pi}\left(\frac{g}{\La^2}\right)^2\,Z^2\,e^2\, m_N
\left\{\, 1- \left(\,1-\frac{2\,M_{\chi N}^2}{m_N^2}\,\right)\frac{m_N \, E_R}{2\, M_{\chi N}^2\,v^2}
\right\} |F_c(E_R)|^2\, \,,
\ee
\end{widetext}
where $E_R$ is the nuclear recoil energy, with a typical value $\sim 1-100$ keV, and
$F_c(E_R)$ is the nuclear form factor, which accounts for the loss of coherent scattering
at large momentum transfer.

The differential scattering rate is then
\begin{equation}
\label{srate}
\frac{dR}{dE_R} = N_T \frac{\rho_\chi}{m_\chi} \int_{v_{min}(E_R)}^{v_{max}} dv f(v) v
\frac{d\sigma}{dE_R},
\end{equation}
where $N_T$ is the number of targets in the detector, $\rho_\chi$ is the local dark matter density, and
$f(v)$ gives the velocity distribution of the dark matter in the lab frame.  The lower limit of
integration is the minimum velocity needed to produce a recoil energy of $E_R$:
\begin{equation}
v_{min} = \left(1 + \frac{m_N}{m_\chi}\right)\sqrt{E_R/2m_N},
\end{equation}
while $v_{max}$ is the maximum dark matter velocity in the lab frame, given by the sum of
the halo escape velocity and the velocity of the earth through the halo.

Eqs. (\ref{dsigma}) and (\ref{srate}) must be convolved with the detector efficiency for any given
experiment to determine the overall detection rate for dark matter.  Here, however, we will take a
simpler approach (loosely modeled on that of Ref. \cite{Nobile}), comparing
the predicted detection rate for anapole dark matter to the detection rate for ``standard" WIMP dark matter with a nuclear
contact interaction (CI) .  This provides additional insight into the behavior of our model
compared to WIMP dark matter, but with the drawback that
the resulting limits on anapole dark matter are less precisely determined.

More specifically, we calculate the ratio
\begin{equation}
{\cal R} = \frac{(dR/dE_R)_{\textrm{anapole}}}{(dR/dE_R)_{\textrm{CI}}},
\end{equation}
where $(dR/dE_R)_{\textrm{anapole}}$ is given by Eqs. (\ref{dsigma})-(\ref{srate}), while
$(dR/dE_R)_{\textrm{CI}}$ is the differential scattering rate for a spin-independent
contact
interaction, derived from
\begin{equation}
\label{CI}
\left(\frac{d\sigma}{dE_R}\right)_{\textrm{CI}} = \frac {A^2 \sigma_n m_N}{2 v^2 M_{\chi n}^2} |F_c(E_R)|^2,
\end{equation}
where $A$ is the total number of nucleons in the target nucleus, $\sigma_n$ is the
dark matter-nucleon cross section, and $M_{\chi n}$ is the reduced mass of the dark matter-nucleon
system.  Eq. (\ref{CI}) assumes that the dark matter particle couples identically to neutrons
and protons.  Generally speaking, Eqs. (\ref{srate}) and (\ref{CI}) go into the standard
calculation for spin-independent dark matter interactions, with the experimental results then
used to place an upper bound on $\sigma_n$.

Using Eqs. ({\ref{dsigma}), (\ref{srate}), and (\ref{CI}), we find that
\begin{widetext}
\begin{equation}
\label{ratio}
{\cal R} = 4 \alpha \left(\frac{g}{\Lambda^2}\right)^2\left(\frac{1}{\sigma_n}\right)
\left(\frac{Z}{A}\right)^2
M_{\chi n}^2 \left[ \frac{I_1}{I_0} + \left(\frac{1}{m_N} - \frac{m_N}{2 M_{\chi N}^2}\right)E_R \right],
\end{equation}
\end{widetext}
where $I_1$ and $I_0$ are integrals over the dark matter velocity distribution given by:
\begin{eqnarray}
I_1 &=& \int_{v_{min(E_R)}}^{v_{max}} dv ~f(v) v,\\
I_0 &=& \int_{v_{min(E_R)}}^{v_{max}} dv ~f(v) \frac{1}{v}.
\end{eqnarray}

To compare the limits on WIMP dark matter to the limits on anapole dark matter,
we select a particular value for $m_\chi$ and use the value for $g/\Lambda^2$ shown in Fig. 4.  For
$\sigma_n$, we take the boundary of the excluded region for a given experiment, and examine the
value of $\cal R$ as $E_R$ ranges over the values of the recoil energy probed by that experiment.
Then, if ${\cal R} < 1$ over this entire range in $E_R$, the given mass $m_\chi$ with anapole interaction
sufficient to account for the dark matter is not ruled out by the experiment under consideration.
On the other hand, if ${\cal R} > 1$ over the entire $E_R$ range, the experimental results do rule out this
value for $m_\chi$.  In the intermediate regime, where ${\cal R}$ changes from
$ < 1$ to $> 1$ over the detectable range in $E_R$, no conclusion can be drawn
without a more detailed calculation.

Eq. (\ref{ratio}) shows the $(Z/A)^2$ dependence typical of comparisons between
electromagnetically-interacting dark matter and dark matter coupling identically to all nucleons.
While $Z/A$ is larger for lighter target nuclei, the current XENON100 225 live day results
\cite{XENON100} are so much more restrictive than any other published limits
over much of the mass range of interest that we will use them here
as our standard of comparison.  Taking $E_R$ to lie in the
range from 6.6 keV to 43.3 keV, and using the truncated Maxwell-Boltzmann velocity
distribution for $f(v)$ in Ref. \cite{Nobile} (ignoring the small
seasonal contribution from the earth's motion around the sun) we find that ${\cal R} < 1$ for all $m_\chi < 80$ GeV,
indicating that this mass range for anapole dark matter cannot be ruled out by the current XENON100
results.  (We have examined different models
for $f(v)$ and included seasonal effects to verify
that this main conclusion is independent of our choice of $f(v)$).
We find that ${\cal R}$ reaches a maximum value $\sim 1/4$ for $m_\chi \sim 30 -
40$ GeV, indicating
that such a particle would lie just below the current threshold for detection by
XENON100. These anapole results contrast sharply with the results for electric or magnetic dipole dark matter,
which allow the lower mass range $m \la 10$ GeV \cite{Sigurdson,Masso,Fortin}, but exclude all masses above
10 GeV.

Although we restricted our freeze-out calculations to
$m_\chi < 80$ GeV,
the additional annihilation channels that open up at $m_\chi \ga 80$ GeV will increase the annihilation
rate at a given value of the anapole moment, so that the correct relic abundance will be achieved for
a smaller value of $g/\Lambda^2$, with no effect on the DM-nucleus cross section.  So we expect masses
in this higher mass range to be even less detectable, and we are justified in concluding that
$m_\chi \ga 80$ GeV is also currently allowed by XENON100.  The main constraint on these heavier
masses comes from the theoretical requirement that $m_\chi < \Lambda$, which is violated for sufficiently
large masses, as is apparent from the calculation in Sec. III.

\section{Discussions and Conclusions}
\label{sec:discuss}

Electromagnetically interacting particles provide a relatively simple model for
dark matter:  once the magnitude of the interaction is fixed to provide the correct
relic abundance for a given mass, there is single unique prediction for the signal
in direct detection experiments at that mass.  (We note in passing
that these models can be considered an example of isospin-violating dark matter proposed by \cite{Feng}).
Unfortunately, for the case of electric or magnetic dipole dark matter, the interaction is sufficiently
strong that direct detection experiments already rule out such models for the entire
range over which such experiments have reasonable sensitivity ($m \ga 10$ GeV).

In contrast, we have shown that a particle interacting exclusively through an anapole
moment cannot currently be excluded at any mass by direct detection experiments,
while such a particle would be close to the XENON100 threshold of detection if
$m_\chi \sim 30-40$ GeV.
(Although $m_\chi \la
5-10$ MeV is excluded by CMB measurements).  Further, the anapole is the only allowed
electromagnetic moment for Majorana dark matter.
While we have not examined collider signatures for this
model, the electromagnetic anapole is clearly an interesting new model for
dark matter and is worthy of more detailed study.

\acknowledgments

We thank D. Hooper and T. Weiler for helpful discussions.
C.M.H. and R.J.S. were supported in part by the Department of Energy (DE-FG05-85ER40226).

\appendix

\section{~~The Cross-Section for $\chi\,\bar{\chi} \rightarrow f\,\bar{f}$}

All of the calculations below will be done with 1/2 times the anapole interaction operator in Eq. \eqref{lagrangian}. This factor of 1/2 is a convention invoked to take care of the self-conjugacy of Majorana fermions. It will cancel out the 2! symmetry factor at any vertex of a Feynman diagram involving a pair of Majorana fermions.

The scattering amplitude for the process $\chi (p)\,\bar{\chi}(p') \rightarrow f (k)\,\bar{f}(k')$ is given by
\bea
\mathcal{M} =\frac{-i \,g\, e}{\La^2}\, \bar{u}_s(k) \,\gamma_\mu \,v_{s'}(k')\,\bar{v}_{r'}(p')\,\gamma^\mu\,\gamma^5\, u_r(p)\,,
\eea
where the subscripts $r, r', s, s'$ are the spin indices of the corresponding fermions. Averaging over the initial polarizations and summing
over the final polarizations, we get
\bea
\overline{|\mathcal{M}|^2} &=& \frac{g^2\,e^2}{4\,\La^4}\,
\textrm{Tr}\,\left\{(\,\,\!\!\not\!k+m_f)\,\gamma_\mu\,(\,\,\!\!\not\!k'-m_f)\,\gamma_\nu\right\} \nonumber \\
&& ~~\textrm{Tr}\,\left\{(\,\,\!\!\not\!p'-m_\chi)\,\gamma^\mu\,\gamma^5\,(\,\,\!\!\not\!p+m_\chi)\,\gamma^\nu\,\gamma^5\right\}\,.
\eea

In the CM frame, we have $p= (E, \vec{p})$, $p'= (E, -\vec{p})$, $k= (E, \vec{k})$ and $k'= (E, -\vec{k})$,
where $E=\sqrt{|\vec{p}|^2+m_\chi^2}$. After some algebra, we obtain
\bea
\overline{|\mathcal{M}|^2} &=& \frac{4\,g^2\,e^2\, E^4}{\La^4}\,v_{\textrm{rel}}^2 \nonumber \\
&& \left\{\left(\,1+\frac{m_f^2}{E^2}\,\right)+\left(\,1-\frac{m_f^2}{E^2}\,\right)\cos^2\theta\right\}\,,
\eea
where $v_{\textrm{rel}}=2\,v$\, with $v=|\vec{p}|/E$ being the velocity of each annihilating dark matter in the CM frame, and $\theta$ is the
angle between $\vec{p}$ and $\vec{k}$.

The differential cross-section is given by
\bea
\frac{d\sigma_{\chi\,\bar{\chi} \rightarrow f\,\bar{f}}}{d\Omega}= \frac{\sqrt{1-\frac{m_f^2}{E^2}}}{v} \frac{1}{64\,\pi^2\,(2\,E)^2}\,\overline{|\mathcal{M}|^2}\,.
\eea
It is then straightforward to calculate the total cross-section
\bea
\sigma_{\chi\,\bar{\chi} \rightarrow f\,\bar{f}}\;v_{\textrm{rel}}= \frac{2\,g^2\,\alpha\,m_\chi^2}{3\,\La^4}\,v_{\textrm{rel}}^2 \,,
%\sigma_{\chi\,\bar{\chi} \rightarrow f\,\bar{f}}\;v_{\textrm{rel}}= \frac{2}{3}\,\left(\frac{g}{\La^2}\right)^2 \alpha\; m_\chi^2 \,,
\eea
where $\alpha=e^2/4\pi \approx 1/137$ is the fine structure constant and we have made the approximation $m_f\ll E\approx m_\chi$.

The thermally averaged relative velocity in the CM frame is given by $\frac12 (\frac12\,m_\chi) \langle\, v_{\textrm{rel}}^2\,\rangle =\frac32\,T$
where $T$ is the temperature. This implies that $\langle\, v_{\textrm{rel}}^2\,\rangle = 6\,T/m_\chi$ and hence
\bea
\langle\,\sigma_{\chi\,\bar{\chi} \rightarrow f\,\bar{f}}\;v_{\textrm{rel}}\,\rangle = \frac{4\,g^2\,\alpha\,m_\chi^2}{\La^4}\left(\frac{T}{m_\chi}\right)\,.
%\langle\,\sigma_{\chi\,\bar{\chi} \rightarrow f\,\bar{f}}\;v_{\textrm{rel}}\,\rangle= 4 \left(\frac{g}{\La^2}\right)^2 \alpha\; m_\chi^2 \,,
\eea \\

\section{~~DM-Nucleus Scattering Cross-Section}

The scattering amplitude for the process $\chi (p)\,N(p') \rightarrow \chi(k)\,N(k')$ is given by
\bea
\mathcal{M} =\frac{-i \,g\,Z\, e}{\La^2}\, \bar{u}_{s'}(k') \,\gamma_\mu \,u_{s}(p')\,\bar{u}_{r'}(k)\,\gamma^\mu\,\gamma^5\, u_r(p)\,,
\eea
where the subscripts $r, r', s, s'$ are the spin indices of the corresponding fermions and $Z$ is the atomic number of the nucleus. Averaging over the initial polarizations and summing
over the final polarizations, we get
\bea
\overline{|\mathcal{M}|^2} &=& \frac{g^2\,Z^2\,e^2}{4\,\La^4}\,
\textrm{Tr}\,\left\{(\,\,\!\!\not\!k'+m_N)\,\gamma_\mu\,(\,\,\!\!\not\!p'+m_N)\,\gamma_\nu\right\} \nonumber \\
&& ~~\textrm{Tr}\,\left\{(\,\,\!\!\not\!k+m_\chi)\,\gamma^\mu\,\gamma^5\,(\,\,\!\!\not\!p+m_\chi)\,\gamma^\nu\,\gamma^5\right\}\,,
\eea
where $m_N$ is the mass of the nucleus.

In the CM frame, we have $p= (E_\chi, \vec{p})$, $p'= (E_N, -\vec{p})$, $k= (E'_\chi, \vec{k})$ and $k'= (E'_N, -\vec{k})$.
After some algebra, we obtain
\bea
\overline{|\mathcal{M}|^2} &=& \frac{8\,g^2\,Z^2\,e^2}{\La^4}\, m_\chi^2\,m_N^2\,v^2\nonumber \\
&&\left\{\left(\,1+\cos\theta\,\right)+\left(\,1-\cos\theta\,\right)\left(\,\frac{2\,M_{\chi N}^2}{m_N^2}\,\right)\right\}\,,
\eea
where $M_{\chi N} = m_\chi\,m_N/(m_\chi+m_N)$ is the reduced mass of the $\chi$-nucleus system. Note that
$|\vec{p}\,|= |\vec{k}|= M_{\chi N}\,v $ where is $v$ is the velocity of $\chi$ in the lab frame.

In the non-relativistic limit,
we have the differential cross-section
\bea
\frac{d\sigma}{d\Omega} \approx \frac{1}{64\,\pi^2\,(\,m_\chi+m_N\,)^2}\,\overline{|\mathcal{M}|^2}\,.
\eea
A more relevant quantity for direct detection is $d\sigma/dE_R = (d\Omega/dE_R) (d\sigma/d\Omega) $ where $E_R$
is the nuclear recoil energy. For small momentum transfers, we have $d\Omega/dE_R = 2\,\pi\,m_N / (M_{\chi
N}^2 \, v^2)$ and so
\bea
\frac{d\sigma }{dE_R} &=& \frac{1}{2\,\pi}\left(\frac{g}{\La^2}\right)^2\,Z^2\,e^2\, m_N \nonumber \\
&& ~~\left\{\,1- \left(\,1-\frac{2\,M_{\chi N}^2}{m_N^2}\,\right)\frac{m_N \, E_R}{2\, M_{\chi N}^2\,v^2}\, \right\}\,.
\eea

Finally, we have the total cross-section
\bea
\sigma =\frac{1}{2\,\pi}\left(\frac{g}{\La^2}\right)^2\,Z^2\,e^2\, M_{\chi N}^2\,v^2
\,\left(\,1+\frac{2\,M_{\chi N}^2}{m_N^2}\,\right)\,.
\eea \\

{}

\end{document}